# Are researchers that collaborate more at the international level top performers? An investigation on the Italian university system[1]

*Giovanni Abramo*[a,b,*], *Ciriaco Andrea D'Angelo*[a] *and Marco Solazzi*[a]

[a] Laboratory for Studies on Research and Technology Transfer
School of Engineering, Dept of Management
University of Rome "Tor Vergata"

[b] National Research Council of Italy

**Abstract**

The practice of collaboration, and particularly international collaboration, is becoming ever more widespread in scientific research, and is likewise receiving greater interest and stimulus from policy-makers. However, the relation between research performance and degree of internationalization at the level of single researchers still presents unresolved questions. The present work, through a bibliometric analysis of the entire Italian university population working in the hard sciences over the period 2001-2005, seeks to answer some of these questions. The results show that the researchers with top performance with respect to their national colleagues are also those who collaborate more abroad, but that the reverse is not always true. Also, interesting differences emerge at the sectorial level. Finally, the effect of the nation involved in the international partnership plays a role that should not be ignored.

# 1. Introduction

Collaboration is a fundamental aspect of scientific research activity. The motives for collaboration are many, however most can probably be attributed to a "pragmatic attitude to collaboration" (Melin, 2000). Indeed, collaboration constitutes an intrinsic feature of scientific research, to the extent that is probably more a need than a choice (Beaver & Rosen, 1978). It is no accident that over time, the resort to collaboration in research has systematically increased (Hicks & Katz, 1996; Wuchty et al., 2007; Schmoch & Schubert, 2008). Collaborations at the international level have also registered a constant increase (Zitt & Bassecoulard, 2004; Archibugi & Coco, 2004). International collaboration assumes a particularly important role, due to the potentials linked to the differences between researchers in scientific and cultural background. It seems reasonable to assume that in international collaborations, precisely because of the differences between partners, the expected results would be greater. Studies of group creativity confirm that it is diversity rather than conformity that lead to more innovative and higher quality results (De Dreu & West, 2001). Researchers from different nations who collaborate together have more probability of learning new (to them) notions, techniques and methodologies from one another, and thus of increasing their personal knowledge assets (Burt, 1992).

However, collaborations also involve transaction costs: needs to negotiate and mediate objectives, choose methodologies, deal with results, manage logistics for communications, manage gatherings and face-to-face meetings, and for further coordination needs (Landry & Amara, 1998). Olson & Olson (2000) have demonstrated that as geographic distance between research partners increases so does the probability of failure and underperformance. In international collaborations, compared to domestic ones, just as one can expect greater results from the heterogeneity of resources (both intellectual and other), one can equally assume the occurrence of greater transaction costs, such as those deriving from cultural and linguistic barriers, or from travel over greater distances. Schmoch & Schubert (2008) underline that "the successful organisation of an international co-operation is more demanding than that of a purely national one".

Overall, the benefits of collaboration are held to outweigh costs and, because of this, collaborations are generally encouraged by the various levels of governance. The literature on analysis of the effects of collaboration for research output indeed demonstrates that scientific output resulting from collaboration has a significantly greater impact compared to that produced from intra-mural collaboration (Wuchty et al., 2007). Specific to international partnerships, Narin & Whitlow (1990) very early demonstrated a significantly favorable differential in impact for publications co-authored by scientists from different nations. In a recent work, Abramo et al. (2010) demonstrate that both productivity and average impact of output are positively correlated to the degree of international collaboration achieved by a scientist. The present work is intended to investigate further on this front, comparing the level of internationalization for scientists of top research performance with the levels for their colleagues, and to also investigate the counter-aspect, meaning analyzing the relation between international collaborations and research performance at the level of the single scientist. The study addresses the following specific research questions:

- Are researchers with the best research performance also those that collaborate more at the international level?

- As a counter inquiry, are researchers that collaborate more at the international level also those that achieve best performance?
- Is there a "partner nation" effect on performance and internationalization?

To attempt a response to these questions, the study will analyze international co-authorships in the publications indexed on the Web of Science (WoS) between 2001 and 2005, produced by Italian university researchers working in the hard sciences.

The next section of the study describes the methodological approach and the indicators used to characterize each researcher in terms of performance and degree of internationalization. Section 3 presents the results concerning the first two research questions. The elaborations responding to the third question are presented in Section 4. The concluding section comments on the results and indicates possible directions in future research.

## 2. Methodology, dataset and indicators

To initiate the response to the research questions posed, Italian university researchers were first subdivided into subsets according to several criteria: i) high scientific performance, ii) high degree of internationalization, iii) presence of collaboration with a specific foreign nation. To attempt to respond to the first two questions the study then examines whether scientists who excel along one of the two dimensions considered (research performance or internationalization) are also characterized, compared to the rest of the population, by higher values along the other dimension. To reply to the third research question, the study examines if Italian researchers who have realized at least one publication in co-authorship with foreign colleagues of a specific nation present levels of research performance and degrees of internationalization that are different from those who have not collaborated exclusively with the same given nation.

The next subsection describes the choices made in characterizing researchers along the two dimensions analyzed (research performance and degree of internationalization of their research activity) and the dataset on which the analysis is based.

### 2.1 Methodological approach

The present work is decidedly bibliometric in character and is based on the co-authorship of publications in international journals. Limiting the analysis to the so-called hard sciences certainly renders the bibliometric approach robust, from the point of view of representation of output. The principle limitations remain that scientific collaborations do not always lead to publication of results and that co-authorship of a publication does not necessarily indicate real collaboration (Melin & Persson, 1996; Katz & Martin, 1997; Laudel, 2002; Vuckovic-Dekic, 2003). Notwithstanding these limitations, investigation of co-authored publication is still one of the most tangible and best documented approaches for the measurement of collaboration in research (Price & Beaver, 1966; Cockburn & Henderson, 1998; McFadyen & Cannella, 2004; Glänzel & Schubert 2004). Objective data such as those based on co-authorship permit the conduct of analysis that is objective, quantifiable, non-invasive and relatively low-cost (Katz & Martin, 1997). In addition, it is possible to use a very high number of observations, not realizable through alternative approaches, such as sample-based surveys, and thus achieve results that are statistically more robust (Smith & Katz, 2000).

**2.2 Data, sources and field of observation**

The data used in this work were obtained by elaboration of the raw data of all WoS publications authored by Italian universities[2] in the period 2001-2005. Each publication was attributed to its Italian university authors through the application of a complex algorithm that identifies the addresses and carries out the disambiguation of the true identity of the authors[3]. Under the current Italian university system, each scientist belongs to a so-called "scientific disciplinary sector" (SDS). Each SDS in turn forms part of a "university disciplinary area" (UDA). The hard sciences, on which this work focuses, are composed of 9 UDAs[4] and 205 SDSs. For a more reliable representation of the phenomena investigated, the analysis is limited to the 165 hard science SDSs in which at least 50% of the member scientists published at least one article in the period under consideration[5]. Further, for greater robustness, the dataset excludes those scientists who entered or left the university system in the period under observation, or who changed SDS or university in this period. From examination of the census provided by the CINECA database of the Italian Ministry of University and Research[6], it results that 26,273 scientists held a stable faculty position over the observed period, in the 165 SDSs. Using the algorithm for identification and disambiguation, it is possible to identify 128,487 publications indexed in the WoS that were produced by these scientists over the five years from 2001 to 2005.

Returning to the discussion of limitations and assumptions, it is significant to note that the observations carried out do not concern a sample, but rather the entire population under analysis: the entire scientific production (censused in the WoS) of Italian university researchers in the 165 SDSs taken into consideration.

**2.3 Indicators**

To characterize each university researcher along the two dimensions studied in this work (research performance and degree of internationalization), we have selected six indicators that attempt to gather various aspects, as defined below.

*Research performance indicators*
The research performance of a scientist is analyzed in consideration of both productivity and average impact of the research product. The first two of the following

---

[2] In Italy there are 95 universities: excluding 11 "distance learning" universities all the others can be considered research universities, regardless of their size.
[3] The algorithm is presented in a manuscript which is currently under consideration for publication in another journal. A short abstract is available at
http://www.disp.uniroma2.it/laboratorioRTT/TESTI/Working%20paper/Giuffrida.pdf
[4] Mathematics and computer sciences; physics; chemistry; earth sciences; biology; medicine; agricultural and veterinary sciences; civil engineering and architecture; industrial and information engineering.
[5] The complete list of the 165 SDS considered is provided on:
http://www.disp.uniroma2.it/laboratorioRTT/TESTI/Indicators/ssd.html
[6] http://cercauniversita.cineca.it/php5/docenti/cerca.php

indicators concern productivity and the third concerns quality/impact[7]:
- Productivity (P): the number of publications authored by a scientist in the period under observation;
- Fractional Productivity (FP): total of the contributions to publications authored by a scientist, the contribution for each publication being considered as the reciprocal of the number of coauthors[8];
- Average Quality (AQ): the mean value of quality of publications authored by a given scientist, the quality of each publication being proxied by citations of that publication divided by the average of citations of all publications of the same type (article or review), in the same year and falling in the same subject category.

The choice of these indicators is based on the assumption that the capacity to attract a foreign partner and his/her willingness to collaborate with an Italian scientist increases with the increasing number of articles realized by the latter (P), and by the quality of these articles (AQ). Since it is also necessary to evaluate intensity of collaboration, it also appears opportune to include an indicator of productivity that takes account of the number of co-authors involved in publications by the scientist (FP).

*Indicators of internationalization*

To analyze the degree of internationalization in the work of a scientist, three sub-dimensions are identified: i) intensity, measured by the number of international publications; ii) propensity, or the tendency to publish with foreign colleagues; iii) amplitude, or number of nations involved in publications authored by a scientist. Each sub-dimension is investigated with a specific indicator (Table 1):
- International Collaboration Intensity (ICI): total of publications authored by a scientist in co-authorship with at least one researcher from a foreign organization (referred to as "cross-national publications");
- International Collaboration Rate (ICR): percent ratio of ICI to total publications by an author[9];
- International Collaboration Amplitude (ICA): total of foreign nations represented in cross-national publications by a given scientist[10].

In order to narrow distortions inherent in measurements performed over different scientific fields (generally characterized by different coverage in journals included in the dataset, different "fertility", intensity of citation, degrees of internationalization, etc.)[11], the analysis is always conducted by individual SDS. This means that absolute values of the indicators are calculated for every scientist of the dataset, and through comparison to the same measurements for the other university researchers of the same national SDS, we then obtain percentile rankings for each individual scientist with respect to his/her national colleagues.

---

[7] We use quality as a synonym of impact, although the dominant view in the bibliometric community is that "impact" is a more proper term, with respect to what citations measure, than "quality". In "quality management" literature quality of a product means conformance to requirements. We believe that the requirement of a research product is to have an impact on scientific or technology progress. That is why we tend to use them interchangeably.

[8] For example, for a scientist authoring 2 articles, the first with other 2 co-authors and the second with 3 other co-authors, FP is $1/3 + 1/4 = 7/12$. In the case of the life sciences, different weights have been given to each co-author according to his/her position in the list and the character of the co-authorship (intra-mural or extra-mural).

[9] ICR has a value only for researchers who achieved at least one publication in the WoS over the five-year period under observation: of the 26,273 scientists who held a stable faculty position over the observed period in the 165 SDSs, 21,504 resulted as authors of a publication in the dataset.

[10] For example, if an Italian researcher produced four papers with authors from the USA, France and Spain, seven papers with authors from France and Germany, and 23 papers with authors from the USA, then his/her value for ICA would be four (because he/she collaborated with colleagues from four different countries: USA, France, Spain and Germany). ICA thus takes a value other than zero only for researchers in the dataset that realized at least one publication with foreign co-authorship: the result is 12,511 such researchers in the dataset.

| Dimension | Sub-dimension | Indicators |
|---|---|---|
| Research performance | Productivity | Product (P) |
| | | Fractional Product (FP) |
| | Average Quality | Average Quality (AQ) |
| Degree of internationalization | Intensity | International Collaboration Intensity (ICI) |
| | Propensity | International Collaboration Rate (ICR) |
| | Amplitude | International Collaboration Amplitude (ICA) |

*Table 1: Classification of indicators used.*

## 3. Analysis of individual top scientists on the basis of performance and internationalization indices

This section presents an investigation of if, and to what measure, scientists who excel along one of the two dimensions considered (research performance or internationalization) are characterized, compared to the rest of the population, by values that are also higher along the other dimension. In fact, it can not be assumed that the best researchers for quantity and/or quality of research product will also be those who collaborate more abroad, or vice versa. This is even more true when considering, within each dimension, the "sub-dimensions" represented by the individual indicators.

For each of the six indicators, the entire population of researchers considered is thus divided into two complementary sub-groups: that of the "top scientists", and the remainder of the population. The top scientists are those who stand out for the specific index considered with respect to the rest of the population of researchers. Specifically, for each of the 165 SDSs considered, the top scientists were identified as those situated in the top 10% of the rankings (by percentile) for the particular indicator considered. Obviously, according to the indicator selected, the precise top 10% identified as top scientists is different. The second sub-group is composed of the remaining population of other scientists with a performance lower than the top 10%.

The next two subsections present a comparison of the top scientists with the rest of the population and measurement of the average differences:
- of internationalization, when the top scientists are identified on the basis of a performance index;
- of research performance, when the top scientists are identified on the basis of an un internationalization index.

---

[11] See Abramo et al. (2008) for details.

## 3.1 Top scientists identified on the basis of a research performance index

The top scientists identified on the basis of each of the research performance indices, when compared to their remaining colleagues, show higher average values for every indicator of internationalization: the values of average difference are all positive (Table 2). For example from the first number (27.12) of Table 2 it results that on average a top scientist identified on the basis of P shows a value for ICI, in terms of percentile ranks, bigger (of 27.12) than the one of his colleagues belonging to the rest of the population. It results that researchers characterized by highest performance, considered from either a quantitative (P, FP) or qualitative (AQ) view, collaborate more abroad than do their colleagues, both in absolute (ICI) and relative (ICR) terms, and also show more extended networks (ICA).

|  |  | Internationalization index | | |
|---|---|---|---|---|
|  |  | ICI* | ICR* | ICA** |
| Performance index for top scientists selection | P | 27.12 | 8.07 | 15.54 |
|  | FP | 23.50 | 5.71 | 13.38 |
|  | AQ | 8.56 | 9.22 | 5.27 |

*Table 2: Average difference in percentile ranks, for each indicator of internationalization, between top scientists identified on the basis of performance indicators and the rest of the population of scientists.*
*\* The analysis for indicators ICI and ICR were carried out for the 21,504 researchers resulting as authors of a publication in the dataset.*
*\*\* Analysis for the indicator ICA was carried out for the 12,511 researchers that realized at least one publication with co-authorship abroad.*

Yet, other than the general trend, differences also emerge according the performance indicator considered. The top scientists for productivity (P) present the greatest average differences for number of cross-national publications (ICI: 27.12), as would be expected, and for number of partner nations (ICA: 15.54). Those identified on the basis of FP show the lowest average difference for tendency to internationalization (ICI: 5.71).

On the other hand, of the top scientists, the ones identified on the basis of average quality of output (AQ) are those that show the lowest values of difference for ICI (8.56) and ICA (5.27), but the highest for ICR (9.22). Compared to the top scientists identified on the basis of P or FP, the researchers characterized by a better quality of output thus produce less and involve lower numbers of nations, but in relative terms, collaborate more with foreign authors.

To detect potential differences at the level of discipline, the preceding analysis was detailed for each of the nine UDAs considered. The top scientists in a UDA are the set of the top scientists from each SDS included in that UDA. There are notable differences at the disciplinary level, for each indicator of performance (Table 3).

Concerning P, all three of the internationalization indices show the lowest values of difference for the Civil engineering and architecture UDA (18.85 for ICI, 4.03 for ICR, 4.52 for ICA). The greatest average difference of ICR is seen for Chemistry (14.2, more than 75.96% of the general aggregate value). The Physics UDA is characterized by the highest values of average difference both for ICI (+41.7% compared to the general data) and for ICA (+62.03%). The Physics and Chemistry UDAs also register similar characteristics when considering FP, rather than P (Table 4).

|  | Internationalization index* | | |
|---|---|---|---|
| UDA | ICI | ICR | ICA |
| Civil engineering and architecture | 18.85 (-30.49%) | 4.03 (-50.06%) | 4.52 (-70.95%) |
| Industrial and information engineering | 19.21 (-29.17%) | 6.18 (-23.42%) | 9.89 (-36.37%) |
| Agricultural and veterinary sciences | 20.34 (-25%) | 4.98 (-38.29%) | 8.81 (-43.28%) |
| Biology | 30.93 (14.05%) | 8.89 (10.16%) | 15.84 (1.9%) |
| Chemistry | 31.7 (16.89%) | 14.2 (75.96%) | 19.5 (25.48%) |
| Earth sciences | 26.21 (-3.36%) | 6.38 (-20.94%) | 15.69 (0.94%) |
| Physics | 38.43 (41.7%) | 8.1 (0.37%) | 25.18 (62.03%) |
| Mathematics and information sciences | 26.59 (-1.95%) | 5.66 (-29.86%) | 11.33 (-27.1%) |
| Medical sciences | 25.99 (-4.17%) | 7.79 (-3.47%) | 13.2 (-15.08%) |
| Total | 27.12 | 8.07 | 15.54 |

*Table 3: Average difference for degree of internationalization (in percentile ranks) between top scientists identified for P and the rest of the population: analysis by disciplinary area.*
\* *Brackets show percentage variations of values for average differences in the UDA when compared to the general aggregate values, indicated in the bottom row.*

|  | Internationalization index* | | |
|---|---|---|---|
| UDA | ICI | ICR | ICA |
| Civil engineering and architecture | 16.29 (-30.68%) | 2.23 (-60.95%) | 3.43 (-74.4%) |
| Industrial and information engineering | 16.56 (-29.53%) | 3.9 (-31.7%) | 8.25 (-38.37%) |
| Agricultural and veterinary sciences | 15.87 (-32.47%) | 2.75 (-51.84%) | 6.66 (-50.26%) |
| Biology | 26.41 (12.38%) | 5.8 (1.58%) | 14.06 (5.09%) |
| Chemistry | 28 (19.15%) | 11.57 (102.63%) | 16.88 (26.15%) |
| Earth sciences | 22.54 (-4.09%) | 4.08 (-28.55%) | 14.4 (7.58%) |
| Physics | 29.92 (27.32%) | 2.08 (-63.57%) | 17.54 (31.04%) |
| Mathematics and information sciences | 21.51 (-8.47%) | 1.91 (-66.55%) | 8.9 (-33.54%) |
| Medical sciences | 24.38 (3.74%) | 7.36 (28.9%) | 12.72 (-4.96%) |
| Total | 23.5 | 5.71 | 13.38 |

*Table 4: Average difference in degree of internationalization (in percentile ranks) between top scientists identified for FP and the rest of the population: analysis by disciplinary area.*
\* *Brackets show percentage variations of values for average differences in the UDA when compared to the general aggregate values, indicated in the bottom row.*

The results for AQ (Table 5) are different: the Biology UDA shows the highest values of differences for all the internationalization indices. Chemistry shows the lowest values of average differences for ICI (5.64) and ICR (2.72), while the lowest average difference for ICA (0.45), is seen, as occurred for P and FP, for Civil engineering and architecture.

|  | Internationalization index* | | |
|---|---|---|---|
| UDA | ICI | ICR | ICA |
| Civil engineering and architecture | 8.22 (-3.97%) | 5.9 (-36.01%) | 0.45 (-91.38%) |
| Industrial and information engineering | 9.3 (8.64%) | 8.2 (-11.06%) | 2.69 (-48.9%) |
| Agricultural and veterinary sciences | 7.91 (-7.59%) | 7.37 (-20.07%) | 5.56 (5.53%) |
| Biology | 12.31 (43.81%) | 14.22 (54.23%) | 7.46 (41.53%) |
| Chemistry | 5.64 (-34.11%) | 2.72 (-70.5%) | 3.09 (-41.36%) |
| Earth sciences | 10.25 (19.74%) | 10.5 (13.88%) | 5.76 (9.31%) |
| Physics | 10.59 (23.71%) | 13.14 (42.52%) | 6.97 (32.31%) |
| Mathematics and information sciences | 6.65 (-22.31%) | 6.4 (-30.59%) | 2.84 (-46.1%) |
| Medical sciences | 7.13 (-16.71%) | 9.64 (4.56%) | 4.53 (-14.11%) |
| Total | 8.56 | 9.22 | 5.27 |

*Table 5: Average difference in degree of internationalization (in percentile ranks) between top scientists identified for AQ and the rest of the population: analysis by disciplinary area.*
\* *Brackets show percentage variations of values for average differences in the UDA when compared to the general aggregate values, indicated in the bottom row.*

## 3.2 Top scientists identified on the basis of an internationalization index

Now we conduct the counter-analysis of that just completed, meaning the top scientists, in each SDS, are identified on the basis of an internationalization index, and their research performance (P, FP, and AQ) is compared to that of their colleagues (Table 6). Unlike the preceding analysis, this test no longer verifies the top scientists as showing greater values than their colleagues for every index considered. The top scientists identified on the basis of ICR show negative average differences for P (-3.48) and FP (-6.92). However, considering average quality of product (AQ), the top scientists for ICR show higher values than their colleagues (average difference is +9.23) Thus, in practice, those who have a greater propensity for international collaboration produce less on average than do their colleagues, but their total product (cross-national and domestic) results as being of higher average quality. Again referring to ICR, it can also be seen that this indicator registers the lowest values of average difference, for all 3 performance indices. The top scientists identified on the basis of ICI and ICA do, however, show positive average differences with respect to their colleagues for every performance index. Those who collaborate more at the international level (in absolute terms), or who have more extended networks, thus show research performance that is better than the remainder of the population. The maximum values of average difference (+37.4 for P, +36.0 for FP, +21.0 for AQ) are all verified for ICI: top scientists identified on the basis of ICI show on average, with respect to their colleagues, values for all 3 indices of performance that are notably higher than those for the rest of the population.

|  |  | Performance index | | |
|---|---|---|---|---|
|  |  | P | FP | AQ |
| Internationalization index for top scientists selection | ICI | 37.42 | 35.97 | 21.02 |
|  | ICR | -3.48 | -6.92 | 9.23 |
|  | ICA | 22.74 | 21.03 | 14.26 |

*Table 6: Average difference in percentile ranks, for each indicator of performance, between top scientists identified on the basis of an internationalization index and the rest of the population.*

Once again the analysis was detailed at the level of the single UDA. Concerning ICI (Table 7) the variations seen at the level of disciplinary area are quite contained. In terms of P, the difference in performance between top scientists (for collaboration intensity) and the remainder of the population varies between +31.46 in Civil engineering and architecture and +43.22 in Physics. Very similar situations are also seen when examining the differences for FP and AQ. In all cases, whatever the extent, the top scientists of all the UDAs show values of performance that are higher than those of their colleagues.

Disciplinary variations are more evident when considering ICR (Table 8) and ICA (Table 9). In all the UDAs, with the sole exception of Chemistry, the performance of top scientists for collaboration propensity (ICR) is seen as lower than that for the rest of the population (Table 8), with very marked differences in Physics, for P (-14.7%) and FP (-21.31%). The situation is the opposite for average quality (AQ): here the difference in performance between top scientists for ICR and the rest of the population is one of advantage for the former, in every UDA, with particularly substantial jumps in Civil engineering and architecture (+40.76% higher than the general aggregate), Agricultural and veterinary sciences (+37.37%) and Medicine (+37.59%).

|  | Performance index* | | |
| --- | --- | --- | --- |
| UDA | P | FP | AQ |
| Civil engineering and architecture | 31.46 (-15.93%) | 31.18 (-13.32%) | 20.72 (-1.44%) |
| Industrial and information engineering | 34.49 (-7.83%) | 31.15 (-13.4%) | 19.69 (-6.32%) |
| Agricultural and veterinary sciences | 32.33 (-13.6%) | 30.2 (-16.04%) | 21.04 (0.08%) |
| Biology | 39.53 (5.64%) | 38.51 (7.06%) | 22.79 (8.43%) |
| Chemistry | 39.89 (6.6%) | 37.3 (3.7%) | 19.32 (-8.1%) |
| Earth sciences | 35.55 (-5%) | 34.14 (-5.09%) | 21.43 (1.94%) |
| Physics | 43.22 (15.5%) | 37.26 (3.59%) | 19.99 (-4.92%) |
| Mathematics and information sciences | 37.69 (0.72%) | 36.23 (0.72%) | 17.77 (-15.47%) |
| Medical sciences | 37.06 (-0.96%) | 38.2 (6.2%) | 22.51 (7.09%) |
| Total | 37.42 | 35.97 | 21.02 |

*Table 7: Average difference in performance, by percentile ranks, between top scientists identified on the basis of ICI and the rest of the population: analysis by disciplinary area.*
*\* Brackets show percentage variations of values for average differences in the UDA when compared to the general aggregate values, indicated in the bottom row.*

|  | Performance index* | | |
| --- | --- | --- | --- |
| UDA | P | FP | AQ |
| Civil engineering and architecture | -0.47 (86.49%) | -6.05 (12.57%) | 12.99 (40.76%) |
| Industrial and information engineering | -0.05 (98.56%) | -4.3 (37.86%) | 11.03 (19.49%) |
| Agricultural and veterinary sciences | -2.5 (28.16%) | -6.91 (0.14%) | 12.7 (37.59%) |
| Biology | -2.41 (30.75%) | -5.41 (21.82%) | 11.96 (29.61%) |
| Chemistry | 3.06 (187.93%) | 0.41 (105.92%) | 4.36 (-52.76%) |
| Earth sciences | -11.71 (-236.49%) | -16.93 (-144.65%) | 0.93 (-89.87%) |
| Physics | -14.7 (-322.41%) | -21.31 (-207.95%) | 3.75 (-59.33%) |
| Mathematics and information sciences | -7.2 (-106.9%) | -13.24 (-91.33%) | 1.55 (-83.24%) |
| Medical sciences | -3.27 (6.03%) | -4.46 (35.55%) | 12.68 (37.37%) |
| Total | -3.48 | -6.92 | 9.23 |

*Table 8: Average difference in performance, by percentile ranks, between top scientists identified on the basis of ICR and the rest of the population: analysis by disciplinary area.*
*\* Brackets show percentage variations of values for average differences in the UDA when compared to the general aggregate values, indicated in the bottom row.*

Considering the top scientists identified on the basis of ICA (Table 9), the Physics area is notable for negative differences in performance relative to the rest of the population, for both P and FP (-13.36% and -12.16% respectively). In all the other areas, scientists characterized by particularly extensive collaborative network show a higher average performance relative to the rest of their colleagues. In every case, in terms of AQ, top scientists for collaboration amplitude have performance that is significantly greater than that for the rest of the population.

|  | Performance index* | | |
| --- | --- | --- | --- |
| UDA | P | FP | AQ |
| Civil engineering and architecture | 51.29 (125.54%) | 41.88 (99.14%) | 7.12 (-50.08%) |
| Industrial and information engineering | 72.93 (220.7%) | 65.18 (209.93%) | 11.47 (-19.54%) |
| Agricultural and veterinary sciences | 67.8 (198.16%) | 60.95 (189.82%) | 12.71 (-10.89%) |
| Biology | 56.91 (150.25%) | 52.03 (147.4%) | 16.24 (13.87%) |
| Chemistry | 64.83 (185.08%) | 63.62 (202.54%) | 14.13 (-0.93%) |
| Earth sciences | 18.12 (-20.3%) | 14.16 (-32.68%) | 13.25 (-7.11%) |
| Physics | -13.36 (-158.77%) | -12.16 (-157.82%) | 15.22 (6.75%) |
| Mathematics and information sciences | 75.41 (231.64%) | 65.37 (210.83%) | 10.41 (-26.99%) |
| Medical sciences | 11.92 (-47.6%) | 9.69 (-53.9%) | 14.93 (4.72%) |
| Total | 22.74 | 21.03 | 14.26 |

*Table 9: Average difference of performance, by percentile ranks, between top scientists identified on the basis of ICA and the rest of the population: analysis by disciplinary area.*

*\* Brackets show percentage variations of values for average differences in the UDA when compared to the general aggregate values, indicated in the bottom row.*

### 4. Performance, internationalization and nationality of foreign partner

In this section, the characterization of the researchers of the dataset refers to the nationality of the foreign institutions to whom their co-authors belong. This is aimed at finding out if and how research performance and degree of internationalization vary with the nation involved in collaboration.

The selection of nations/continents for analysis was based on the numbers of cross-national publications by Italian university researchers. In an attempt to detect the most possible aspects of the phenomenon under observation, the selection includes the top four nations for frequency of publication by the population of Italian university researchers over the period under consideration (USA, France, Germany, UK), with an additional selection of four nations and regions considered of emerging importance (China, India, Latin America[12] and Africa) (Table 10).

Over the period considered, Italian researchers realized more publications with the USA than with any other nation. Of the 41,445 publications produced in international co-authorship by the researchers under observation, a full 12,560 of these products (30.3%) were realized with American researchers. The Italian researchers involved in these publications numbered 6,167, or 28.68% of the total considered, and represented all of the 165 SDSs considered.

To attempt to isolate and evidence potential specificities of the nations under observation, we concentrate on the set of Italian researchers that collaborated exclusively with the foreign nation considered. For example, of the 6,167 Italian researchers that collaborated with the USA, there are 2,500 that collaborated only with the USA and not with any of the other 7 nations/regions analyzed.[13] These 2500 are thus compared to the rest of the population, composed of all remaining Italian scientists who collaborated with foreign nations.

|  | Publications | | Italian university researchers | | |
|---|---|---|---|---|---|
| Country | Total publications | Incidence (%) in total Italian cross-national publications | Collaborating with a partner in the country | Incidence (%) in total of university researchers | Number of SDSs |
| USA | 12,560 | 30.3 | 6,167 | 28.68 | 165 |
| France | 6,646 | 16.0 | 3,725 | 17.32 | 160 |
| Germany | 5,831 | 14.1 | 3,376 | 15.70 | 150 |
| UK | 5,772 | 13.9 | 3,229 | 15.02 | 157 |
| Latin America | 1,542 | 3.7 | 998 | 4.64 | 134 |
| Africa | 711 | 1.7 | 531 | 2.47 | 114 |
| China | 555 | 1.3 | 413 | 1.92 | 93 |
| India | 349 | 0.8 | 255 | 1.19 | 64 |

*Table 10: Publications co-authored by Italian university researchers and colleagues of foreign nations considered: 2001-2005.*

---

[12] Under "Latin America" the following countries only were considered: Argentina, Brazil, Chile, Mexico.
[13] Thus these 2,500 researchers also include those who collaborated with countries other than the additional seven considered in this study, during the five year period under consideration.

In general, the researchers who collaborated with each of the foreign nations considered show average values that are higher than their colleagues for two indicators of productivity (P and FP, Table 11). Both the maximum values for differences (7.60 for P and 13.57 for FP) occur for the set of researchers who collaborated with Indian colleagues. The situation is more heterogeneous when examining average quality of research (AQ): scientists who collaborated with the USA show the greatest performance difference (+6.64), while those who collaborated with Africa, China or India show lower performance compared to their colleagues, with respective negative differences of -3.02, -0.73 and -0.32.

| Country | Number of Italian researchers considered | Performance index | | | Internationalization index | | |
|---|---|---|---|---|---|---|---|
| | | P | FP | AQ | ICI | ICR | ICA |
| USA | 2,500 | 7.49 | 6.45 | 6.64 | -2.52 | -0.67 | -5.65 |
| France | 931 | 4.17 | 3.52 | 2.51 | -6.28 | -3.84 | -6.43 |
| Germany | 654 | 1.14 | 1.10 | 1.34 | -7.95 | -3.83 | -7.21 |
| UK | 765 | 2.93 | 2.11 | 2.95 | -6.18 | -3.12 | -6.39 |
| Latin America | 197 | 3.87 | 3.95 | 1.91 | -2.92 | 1.63 | -3.39 |
| Africa | 120 | 2.08 | 0.54 | -3.02 | -1.90 | 1.47 | -2.25 |
| China | 68 | 5.51 | 0.90 | -0.73 | -3.62 | -1.44 | -4.79 |
| India | 39 | 7.60 | 13.57 | -0.32 | -9.11 | -10.69 | -8.58 |

*Table 11: Average difference in percentile ranks for each indicator of performance and of internationalization between scientists who exclusively collaborated with a specific nation and the rest of the population.*

However, considering the indicators of internationalization, the average differences for ICI are all negative (ranging from -9.11 for Indian to -1.90 to Africa), as we would have expected: in effect, the rest of the population, from which the differences are calculated, is composed of researchers who collaborated with more than one of the nations considered or with other foreign nations, and therefore their scientific production in foreign co-authorship is likely greater. The average differences for ICR are also negative values, with the exception of values relative to Latin America (+1.63%) and Africa (+1.47%). For ICA, the values are all negative and range from a low for India (-8.58%) to a high for Africa (-2.25%).

## 5. Discussion and conclusions

This work, taking a bibliometric approach, examines the relationship between research performance and the degree of internationalization of scientific activity, conducting the examination at the level of the individual researcher. The intent was to verify if scientists with the best research performance are also those who collaborate more internationally, and vice versa. To obtain robust results, the dataset selected was unique for its size and completeness, including 124,000 WoS-listed publications over the period 2001-2005, authored by all the Italian university community, made up of more than 26,000 scientists of the hard science disciplines.

The results seem to confirm the initial hypothesis that researchers characterized by the best research performance also have a greater intensity of and propensity towards international collaboration. This applies to both sub-dimensions identified to study the research performance: productivity and average quality of the research products.

In fact, both more productive scientists and the ones with top impact results collaborate more abroad than do their colleagues, both in absolute (number of cross-national publications) and in relative terms (ratio of cross-national publications to total authored publications), and present more extensive collaboration networks (number of different nations represented in cross-national publications).

Although there are notable differences at the disciplinary level, for each indicator of performance, this is especially evident for the top-performer scientists in three basic sciences: chemistry, physics and biology. In these disciplines in fact the scientists characterized by better research performance tend to collaborate internationally more. This is particularly true in chemistry and physics when we consider research productivity, and in biology and physics as far as research quality is concerned. On the other hand, the top scientists in Civil engineering and architecture, regardless of the performance index used to select them, show lower values in terms of international collaboration networks than their colleagues. In this specific discipline the researchers with better research performance collaborate with fewer foreign countries.

Considering the top scientists for international collaboration intensity, the results are a mirror image: these scientists have research performance that is clearly superior to the rest of the population along all aspects considered (quantitative, contributive, qualitative). The same occurs for the top scientists for amplitude of international collaboration. In other words the researchers who author more cross-national papers in absolute terms or with a greater number of foreign countries tend to outperform their colleagues in terms of research outcome (along all the three performance dimensions).

However, when identified for propensity to collaborate abroad, there is no indication that the top scientists achieve superior productivity than their colleagues, except in average impact. These researchers produce less than their colleagues, but on average their research products are of higher quality. As these researchers tend to produce more in relative terms with foreign colleagues, their lower productivity be caused by the higher transaction costs typically involved in international collaborations. The registered greater impact appears completely in line with what has already been noted in the literature: works in international co-authorship on average receive greater citations. This means that those researchers with a higher incidence of cross-national publications among their total publications then register an average impact that is generally superior to that of their colleagues.

Also here there are disciplinary variations: while they are enough contained when the selection criteria for top scientists is based on international collaboration intensity, they are larger for the other two international indices. Again the largest differences are registered in physics and chemistry.

Elaborating the data by nationality of the foreign partner with which the Italian researchers collaborated, tests demonstrate that the greatest difference in productivity with respect to the rest of the population occurs for scientists who collaborate with the USA and India, while those who collaborate with the USA also achieve the highest difference in average impact.

In a context of increasing interest on the part of the policy-maker for "internationalization of research", and so mirrored by interest among management in universities and public research institutions, it seems evident that incentive schemes in favor of foreign collaboration should not substitute, but at most integrate those directed towards stimulating increased performance. This is because, while performance appears directly correlated to intensity/propensity for international collaboration, the reverse correlation is not equally evident.